**Trends in Planetary Science research in the Puna and Atacama desert regions: under-representation of local scientific institutions?**

A. Tavernier[1*], G.A. Pinto[2,3], M. Valenzuela[4,5,6], A. Garcia[1], C. Ulloa[1], R. Oses[7] & B.H. Foing[2,8,9,10]

[1]Instituto de Investigaciones Científicas y Tecnológicas, IDICTEC, Laboratorio de Investigacion de la Criosfera y Aguas, Universidad de Atacama, UDA, Copiapó, Chile.

[2]Instituto de Investigación en Astronomía y Ciencias Planetarias, INCT, Universidad de Atacama, UDA, Copiapó, Chile.

[3]Centre de Recherches Pétrographiques et Géochimiques, CRPG, Université de Lorraine, Nancy, France.

[4]Departamento de Ciencias Geológicas, Universidad Católica del Norte, UCN, Antofagasta, Chile.

[5]Millennium Institute of Astrophysics, MAS, Chile.

[6]Center for Excellence in Astrophysics and associated Technologies, CATA, Santiago, Chile

[7]Centro Regional de Investigacion y Desarrollo Sustentable de Atacama, CRIDESAT, Universidad de Atacama, UDA, Copiapó, Chile.

[8]International Lunar Exploration Working group, ILEWG, EuroMoonMars.

[9]Vrije Universiteit Amsterdam, VUA, Amsterdam, Netherlands.

[10]Universiteit Leiden, Leiden, Netherlands.

* corresponding author : adrien.tavernier@uda.cl

**Abstract**

In 2019 while launching a multidisciplinary research project aimed at developing the Puna de Atacama region as a natural laboratory, investigators within the University of Atacama (Chile) conducted a bibliographic search identifying previously studied geographical points of the region and of potential interest for planetary science and astrobiology research. This preliminary work highlighted a significant absence in foreign publications consideration of local institutional involvement. In light of this, a follow-up study was carried out to confirm or refute these first impressions, by comparing the search in two bibliographic databases: Web of Science and Scopus. The results show that almost 60% of the publications based directly on data from the Puna, the Altiplano or the Atacama Desert with objectives related to planetary science or astrobiology do not include any local institutional partner (Argentina, Bolivia, Chile and Peru). Indeed, and beyond the ethical questioning of international collaborations, Latin-American planetary science deserve a strategic structuring, networking, as well as a road map at a national and continental scale, not only to enhance research, development and innovation but also to protect an exceptional natural heritage sampling extreme environmental niches on Earth. Examples of successful international collaborations such as the field of meteorites, terrestrial analogues and space exploration in Chile or astrobiology in Mexico are given as illustrations and possible directions to follow in order to develop planetary sciences in South America.

**Keywords**

Atacama desert / Bibliographical survey / International collaboration / Planetary science / Puna / South America

**Highlights**

- Significant under-representation of local institutions in the planetary science's field

- Contrast with the field of geoscience, for which local institutions are predominant

- Planetary science studies using the Puna and the Atacama Desert focus mainly on Mars

- Need to generalize local initiatives in planetary science at the South American scale

**1. Introduction**

After mankind's planned return to the Moon within the next decade, the next step in conquering space should lead to manned missions to Mars (Reneau, 2021). In fact, since the Apollo program launched in 1961, numerous missions have attempted to simulate the conditions that astronauts could face in specific terrestrial environments close to those of space (i.e. weightlessness) (Pandiarajan and Hargens, 2020) or other celestial bodies (e.g., Moon, Mars, Titan and Icy moons) (Léveillé, 2010; Groemer and Ozdemir, 2020). Similarly, rovers and other related instruments sent to the Martian surface have also benefited from life-size tests in terrestrial landscapes similar to Mars (Rull et al., 2021; Whittaker et al., 1997). This interest in terrestrial analogs has also echoed in the development of comparative planetology (Farr, 2004) and, more recently in the rise of astrobiology (Martins et al., 2017). More specifically, certain extreme conditions concerning Earth's habitability may present parallels to the chemicophysical conditions encountered on other bodies within the Solar System and thus provide clues to the presence and viability of lifeforms other than on Earth (Preston and Dartnell, 2014).

*1.1. Atacama desert and Puna ecoregions*

The Atacama Desert and its surrounding Puna ecoregions, part of the Central Andes (see Fig. 1) present special features linked to their arid or semi-arid characteristics. Indeed, this aridity has been stable over several millions of years in recent geological times (Dunai et al., 2005; Clarke, 2006). The presence of marked wind erosion (De Silva et al., 2013; Perkins et al., 2019), intermittent rivers (Houston, 2006), vast areas of evaporites in an endoreic context (Marazuela et al., 2019), complex groundwater recharge (Salas et al., 2016), glacial and periglacial geomorphological landforms (García et al., 2017), intense radiation (Matteucci et al., 2012) and the proven presence of extremophiles (Albaraccín et al., 2015) make this volcanic region (De Silva, 1989; Allmendinger et al., 1997) a possible relevant analog for the conditions that Mars may have experienced during its past geological history making this particular place of the Earth one of the hypothetical "hot-spot" for the development of planetary science (De Silva et al., 2002).

*1.2. A definition of the planetary science field*

In reference to this article, planetary science is defined as the temporally and spatially resolved geological, geophysical and geochemical study of celestial bodies orbiting a star. The recent advent of astrobiology also makes it possible to add biology and biochemistry to the three previously mentioned "geo" disciplines. This geoscience-oriented definition is necessarily subjective and therefore excludes from the field of planetary science, the study of exoplanets which is most often linked to studies of stellar characteristics. However following (Unterborn et al., 2020) establishing a close link between these two disciplines seems essential to the authors of this study.

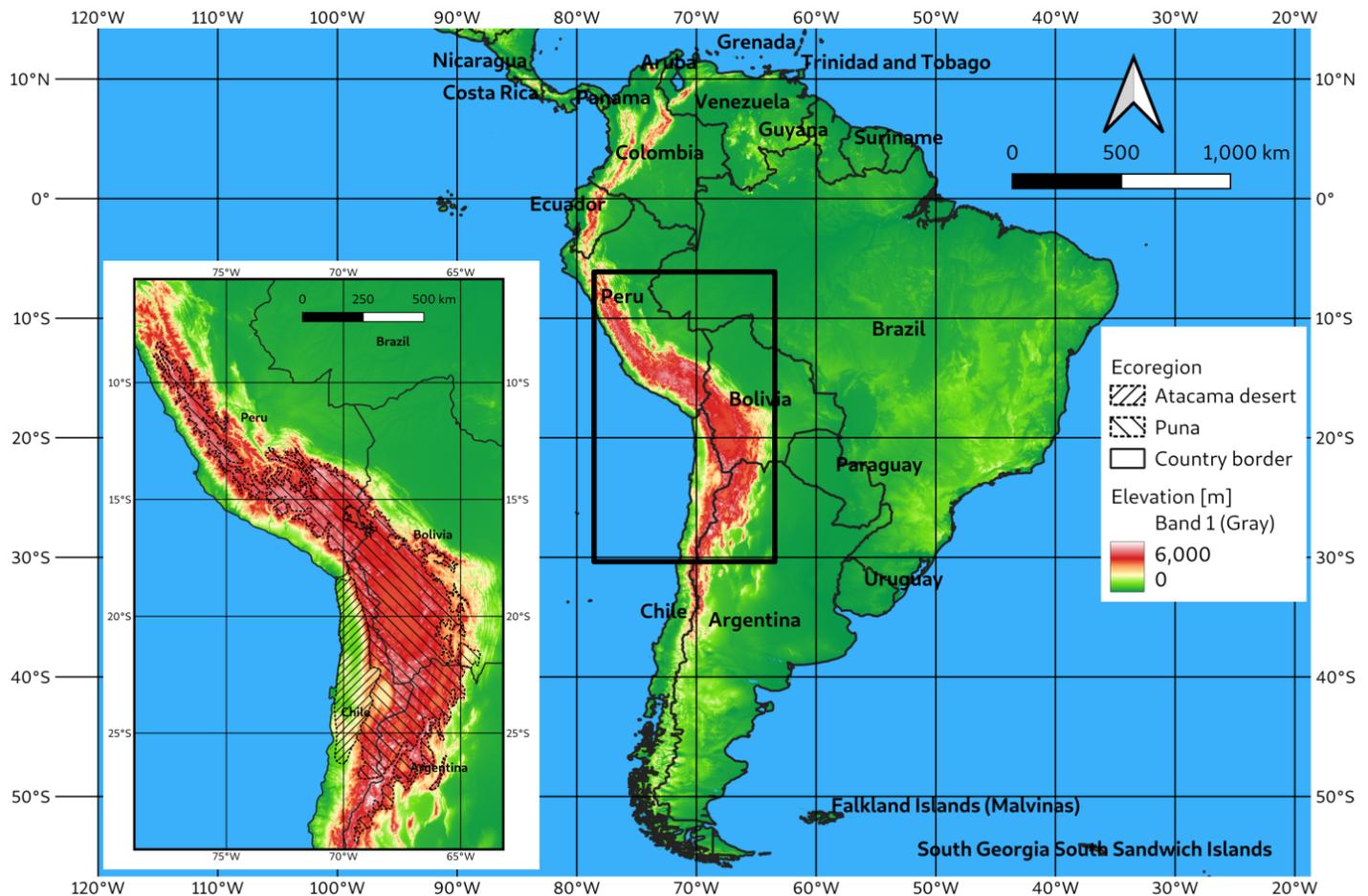

**Fig. 1: Geographical location of the Atacama Desert and the Puna / Digital Elevation Modeling, Global Multi Resolution Terrain Elevation Data 2010 (Resolution: 30 arc-seconds )**

*1.3. Foreign leadership: reality or impression?*

Concerning the use of the Puna and the Atacama Desert as a planetary analog, preliminary bibliographical surveys suggested foreign leadership (i.e. not affiliated to a research organization from Argentina, Bolivia, Chile or Peru). These first impressions were based in particular on several publications concerning the Ojos del Salado region, the world's highest active volcano (6893 meters above sea level | 27°11S / 69°54W), and its hypothetical

potential as a Martian analog (Kereszturi et al., 2020). This foreign leadership could initially be surprising, due to the fact that technical and scientific expertise in geoscience already seems to be present within Argentina, Bolivia, Chile and Peru notably through a history of mining in the central Andes (Alimonda, 2015; Brown, 2012).

By analyzing the origin of scientific publications, this study attempts to confirm or deny this impression of the preponderance of foreign contributions, in planetary science, and resulting directly from the use of the environmental characteristics of the Atacama Desert and/or the Puna regions. The materials and methods section present the methodology used to select and validate the publications included in the analysis. The results section quantifies, through the selected publications, the work contribution of regional and foreign researchers, first in the field of geoscience and then more specifically in planetary science, both involving studies made in the Atacama Desert or in the Central Andes. Eventually, the discussion section returns to the need to improve and develop regional and international expertise and networking to develop a dynamic, ambitious, and efficient planetary science component in South America, taking advantage of the exceptional environmental characteristics of the Atacama Desert and Puna regions. Prospects for the initiation of a ChileMoonMars pole, South American counterpart of the EuroMoonMars program (EMM), part of the International Lunar Exploration Working Group (ILEWG) and dedicated mainly to involve students and young professionals in themes related to space exploration (Foing, 2020) as well as the meteorite field development in Chile will be discussed as possible examples to be followed and generalized.

**2. Materials and Methods**

The search and selection of publications were carried out via the scientific information platform Web of Science (WoS) (Li et al., 2018). Two article selections were elaborated: a geographical selection of studies highlighting the region's natural heritage under consideration (Selection 1) and, included in this first selection, a planetary science selection (Selection 2). In order to avoid bias related to the use of a specific database (Mongeon and Paul-Hus, 2016) and in order to estimate the representativeness of each of the two selections considered , both of them were compared, for strictly identical search criteria, with the Scopus database (Burnham, 2006). The use of these two databases is explained by their multidisciplinary nature, the power and ease of analysis of the results and their restriction to academic publications only. An overlay of publications was also carried out with the Astrophysics Data System (ADS) database (Kurtz et al., 2000) to compare with a platform which is more specialized in planetary science than WoS or Scopus.

*2.1. General criteria for selection and comparison of the different databases*

The period under review considers all the publications in each of the two scientific information platforms with an upper threshold year of 2021 (inclusive), the lower threshold being defined by the date of the first publications available on the different platforms (year 1975 for WoS and ADS, up to year 1788 for Scopus). Proceedings and conference abstracts were excluded. Thus, only peer-reviewed scientific articles in English were considered. After comparison and removal of duplicates, publication recovery rate of both scientific information platforms reached over 70% (74% for Selection 1 and 76% for Selection 2, considering the title of publication and DOI for comparison), which was considered to be satisfactory and justified, for the continuation of the study, the exclusive use of the WoS database. WoS database was favored because of its many articles available (5369 articles to 5288) but also according to better possibilities of statistical analysis of publications (although the results of the analysis for Scopus, similar to those of WoS, are also available in the Supplementary Materials section: Fig. S1, S2, S3, S4 and Table S5).

The comparison with the ADS database is much more nuanced. Indeed, the number of articles available via ADS for Selection 1 is more than three times less than for Scopus or WoS (1684 articles to 5288 and 5369 respectively). Nevertheless, there is an overlap rate of more than 80% with respect to the two generalist databases (82% with WoS and 90% with Scopus). For selection 2, the number of articles is relatively similar for the 3 databases (186 articles for ADS, 165 articles for Scopus and 161 articles for WoS) but with overlap rates between ADS and the two other databases of less than 50% (48% with WoS and 47% with Scopus). These figures tend to confirm the more, even too specialized character of a database like ADS and led to the preferential use of WoS (and Scopus) for this study. The general criteria for the constitution of article selections are summarized in Table 1 and the search results for the different selections, for both WoS and Scopus, in Table 2.

| Language | English |
|---|---|
| Kind of publication | Peer-reviewed articles (conference abstracts and proceedings excluded) |
| Year | Oldest publication – 2021 (included) |
| Geographic | Keywords related to the region under consideration |
| Astronomy | Keywords in order to exclude articles dealing with a telescope or an astronomy-related instrument including the word Atacama (see Table 3) |
| Planetary Science | Keywords in order to select Planetary Science studies (see Table 4) |

*Table 1: Selection criterion considered for this study*

| | Number of articles (excluding duplicates) | | |
|---|---|---|---|
| | Web of Science (WoS) | Scopus | Shared articles |
| Selection 1 | 5369 | 5288 | 3890 (74%) |
| Selection 2 | 161 | 165 | 123 (76%) |

*Table 2: Search results for the WoS & Scopus database and related shared articles*

*2.2. Keywords selection strategy*

Selection 1 was based on the presence in the title, abstract or keywords of a given publication of the terms Atacama, Central Andes, Altiplano or Puna associated with the different naming of the countries under consideration (Argentina, Bolivia, Chile and Peru) and considered as territorial countries or Puna's countries (in contrast to external countries). It is therefore a geographical criterion that has been favored in this study, which focuses mainly on the potential of the Atacama Desert and the arid Central Andes in terms of planetary science and astrobiology research. Both regions have exceptional atmospheric characteristics for astronomical observation, which explains the presence of numerous international observatories in Chile (Torres, 2004). This astronomical activity constitutes an important bias for this study, especially in the appearance of the term "Atacama" in the names of numerous telescopes or astronomical instruments. In fact, while an astronomical site study highlights the regional natural heritage through a detailed analysis of local atmospheric conditions, the same cannot be said of all the scientific studies directly resulting from astronomical observations made by the telescopes present in the region under consideration. Then, a filtering process using keywords was carried out for articles or abstracts quoting a telescope containing the word "Atacama". The need to avoid all publications related to observational astronomy is not insignificant and reflects an important scientific activity in the region. The list of keywords used to obtain Selection 1 is presented in Table 3.

| | Inclusive Geographical Criteria | | Exclusive Astronomical Criteria |
|---|---|---|---|
| | Region | Country | |
| **Keywords** | *atacama / central andes / puna / altiplano* | *chilean / chile / peru / peruvian / bolivia / bolivian / argentina / argentinean / argentinian* | *telescope / cosmic microwave / eso paranal / alma / apex / atacama large millimeter / atacama pathfinder / atacama cosmology / atacama compact / atacama submillimeter / atacama large / atacama large millimetre / atacama submillimetre / chajnantor / sma / socaire* |

*Table 3: Keywords selected to get refereed publications that highlight the natural heritage of the Atacama Desert and the Central Andes (i.e. Selection 1)*

As a result of the regional geographical selection and the exclusive astronomical filter, a first selection was therefore made of works highlighting the regional natural heritage (Selection 1). The second stage consisted in applying a new filter to select the planetary science study of the geographical area under consideration (Selection 2). The keywords used in the first selection are therefore reused, this second selection being a sub-selection of the first one. The constitution of a list of keywords inherent to the field of planetary science was challenging and probably reflects a lack of structuring of the discipline, also linked to its very great disciplinary diversity. By way of comparison, astronomy benefits from a much more marked structuring, illustrated by the preponderance of the International Astronomical Union (IAU) which, for example, has its own list of keywords to standardize its publications.

It was decided to refer to the list of keywords used by the research and analysis programs of the National Aeronautics and Space Administration (NASA) Planetary Science Division ("National Academies of Sciences, Engineering, and Medicine", 2017). The selection criteria to compile a list that could be used in this study are presented in Fig. 2: terms with astrophysical connotations have been removed, as have overly generic words that could be used in a context totally different from that of planetary science, and finally the term regolith has been preferred to dust.

From Selection 1 and Selection 2, filters were used, via the WoS analysis processes, to separate the contributions of the territorial countries (Argentina, Bolivia, Chile and Peru) from the external ones. From a methodological point of view, the data were extracted from the WoS (and Scopus) site and then processed locally with Python codes.

| **Keywords from the Research and Analysis Programs of NASA's Planetary Sciences Division** | |
|---|---|
| **Task:** | ~~Theoretical / Computational~~, ~~Support~~, ~~Sample Analysis~~, ~~Purchase of Major Equipment~~, ~~New observations~~, ~~Mission data analysis~~, ~~Instrument / Technological Development~~, ~~Field-based~~, ~~Experimental~~, ~~Education / Public outreach~~, ~~Archiving / Data restoration~~, ~~Analysis of ground-based data~~, ~~Analog study~~ |
| **Target:** | Venus, ~~Mercury~~ (*confusion with the chemical element*), Martian ~~system~~, ~~Extra-solar~~ ~~planets~~, ~~Earth~~ / Moon ~~system~~, Early solar system, ~~Early Earth~~, ~~Small Bodies~~, Outer Planets, Uranian ~~system~~, Saturnian ~~system~~, Icy bodies, Neptunian ~~system~~, Jovian ~~system~~, ~~Potentially hazardous objects~~, Plutonian ~~system~~, ~~Near-earth objects~~, Meteorites, Kuiper belt ~~objects~~, Trans-neptunian ~~object~~, ~~Comet~~ (*confusion with the Comet Assay methodology*), Asteroids, ~~Solar wind~~, ~~Protoplanetary disks~~, Presolar ~~nebula~~, ~~Non specific planets~~, ~~Interstellar grains~~, ~~Hypervelocity impacts~~, ~~Dust~~ Regolith, Deimos, Phobos, Mars |
| *Remark : Saturn (→ Jovian & Saturnian), Lunar (→ Moon), Neptune (→ Neptunian), Pluton (→ Plutonian), Uranus (→ Uruanian) and Venusian (→ Venus) have been tested but do not add any relevant articles to the Planetary Science selection (Selection 2) and sometimes irrelevant articles (for example Pluton, confusion with geology)* | |
| **Science Discipline:** | ~~Spectroscopy~~, ~~Solar systems dynamics~~, ~~Planetary defense~~, ~~Planetary protection~~, ~~Planetary dynamics~~, ~~Mineral physics~~, ~~Magnetospheres~~, ~~Geophysics~~, ~~Geology~~, ~~Geochemistry~~, ~~Exosphere~~, Cosmochemistry, ~~Atmospheres~~, ~~Astrophysics~~, ~~Astronomy~~, Astrobiology |
| ~~Planetary~~ : too general or too generic<br>~~Planetary~~ : too much astrophysics oriented<br>Planetary : accepted | |
| **Final keyword list for planetary sciences characterization:** | Asteroid, Astrobiology, Cosmochemistry, Deimos, "Early solar system", "Icy body", Jovian, "Kuiper belt", Mars, Martian, Meteorite, Moon, Neptunian, "Outer planet", Phobos, Plutonian, "Presolar nebula", Regolith, Saturnian, Trans-neptunian, Uranian, Venus |

*Table 4: Keywords selected for Selection 2*

## 3. Results

The main results analyzed for WoS are presented in Fig. 2 & 3 (same results for Scopus which show similar trend are available in the section Supplementary Material in Fig. S1 & S2). Two criteria were used to compare territorial (i.e. Puna's countries as Argentina, Bolivia, Chile and Peru) and external countries: the proportion of publications for a given country (bar chart) and the fields of study concerned (pie chart). It is important to note that the analyses performed for the percentage of publications for a given country (in the sense of the institutions involved, not in the sense of the nationality of the authors), as well as the fields of study involved, are non-exclusive. This means that a given publication may involve several different countries and/or fields of study (the sum of the contributions in percentage is therefore always greater than unity). As a reminder Selection 1 corresponds to the geographical selection highlighting the natural heritage of the region of interest, while Selection 2 focuses solely, within this Selection 1, on works in the field of planetary science.

*3.1. Research in geoscience active and locally involved, much more nuanced statement in planetary science*

The Selection 1 (Fig. 2) presents a balanced proportion between the external and the territorial countries on relatively similar study themes involving mainly works in geoscience (i.e. geology, geophysics, and geochemistry), environmental science and ecology. Considering all the contributions on a global scale (Table 5), 64% of the publications involve at least one of the territorial countries, which means, by symmetry, that almost 36% of the articles published did not benefit from any local collaboration.

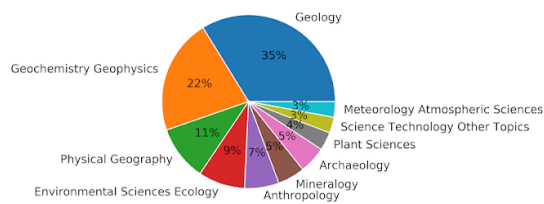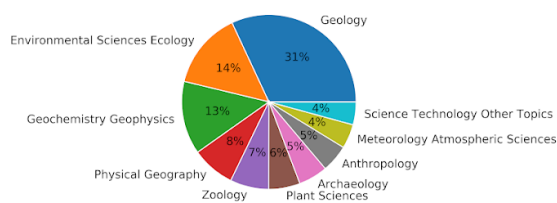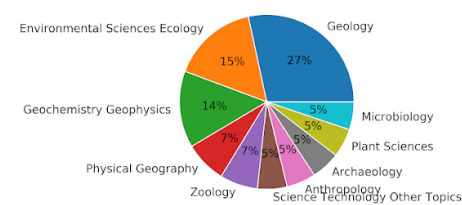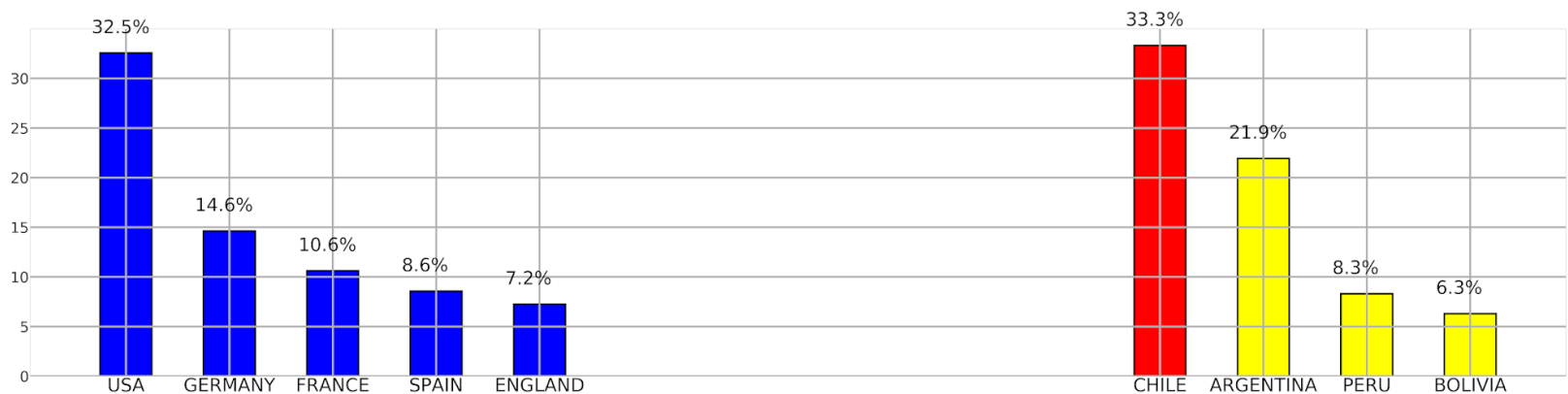

*Fig. 2 WoS data: Relative contribution (according to the total number of publications) to the natural heritage of the Puna Region, in terms of publications, of different countries or group of countries (Selection 1, 5369 articles in total, excluding duplicates) | From top to bottom: Distribution of field of research / percentage of contribution per country*

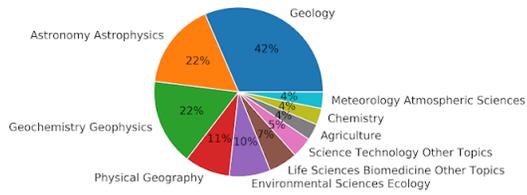
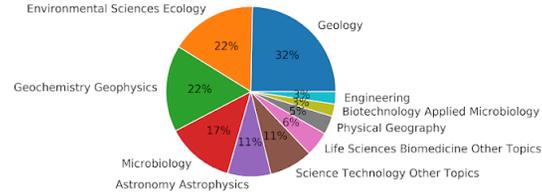
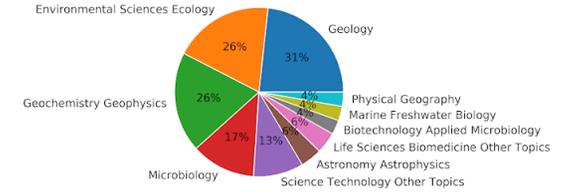
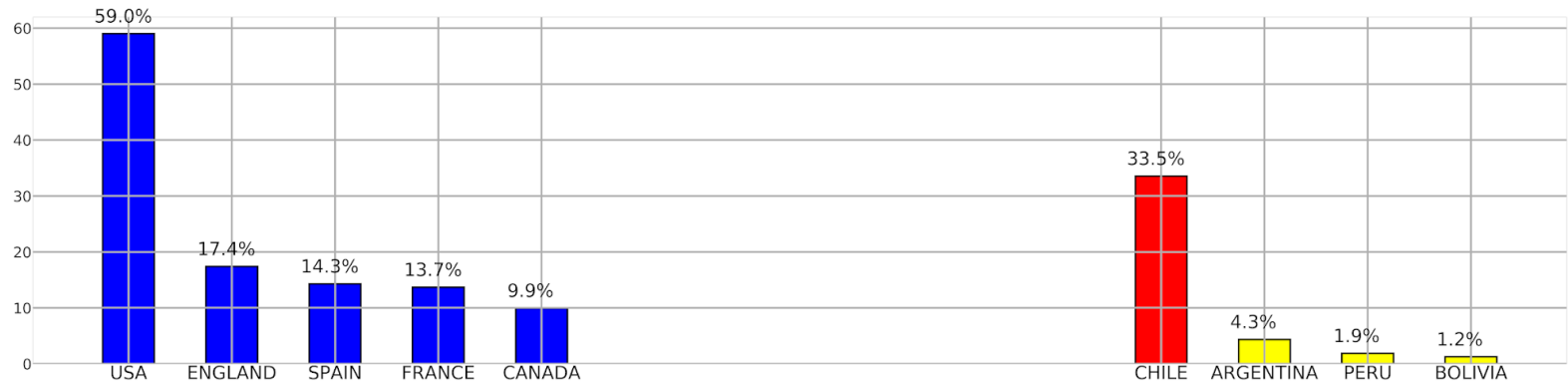

*Fig. 3 WoS data: Relative contribution (according to the total number of publications) to the natural heritage of the Puna Region, in terms of publications in Planetary Science, of different countries or group of countries (Selection 2, 161 articles in total, excluding duplicates) | From top to bottom: distribution of field of research / percentage of contribution per country*

The Selection 2 of articles shows a marked leadership of the USA, a historical actor in the development of planetary science and the use of terrestrial analogs for space exploration (Lofgren et al., 2011; Messeri, 2014). The contribution of Chile remains, in proportion (around 33%), almost identical to that of Selection 1, which testifies to proven research activity in the field of planetary science. For the other territorial countries (i.e. Argentina, Bolivia, and Peru), the contribution is marginal and decreases sharply compared to Selection 1. Considering all the contributions on a global scale (Table 5) another meaningful change with respect to Selection 1 is the significant decrease in the contribution of territorial countries (from 64% to 40%), thus confirming the foreign leadership and the lack of involvement of local partners in the valorization of scientific research carried out in the Atacama Desert and Puna's ecoregions in the field of planetary science and astrobiology.

|  | Number of articles (excluding duplicates) | | |
| --- | --- | --- | --- |
|  | **Total** | **External countries only** | **Participation of at least one territorial country** |
| **Selection 1** | 5369 | 1957 (36%) | 3412 (64%) |
| **Selection 2** | 161 | 96 (60%) | 65 (40%) |

*Table 5: Overall contribution to Selection 1 & 2 from external and territorial countries (WoS data)*

*3.2. The importance of specialization in planetary science*

It is interesting to note in Fig. 3 in the pies related to Selection 2 the appearance, for the external countries, of the field "Astronomy and Astrophysics" as a major topic of study, this despite the use of keywords supposed to eliminate works related to the use of telescopes and astronomical-related instruments present in the geographical region considered. This probably means that part of the works anchored in the Puna region or in the Atacama Desert and conducted by external institutions has an astronomical connotation, in the sense of studies applied to the understanding of bodies other than the Earth. This suggests studies that are thought out and structured with a planetary science purpose, and therefore with truly specialized teams in this field. On the other hand, the main fields of research involved in Selection 2, for Chile as well as for the territorial countries (whose distribution is dominated by Chile), do not present major changes with respect to Selection 1. This is interpreted as research in planetary science that would be, for the most part, a research qualified as one of opportunities (i.e. studies carried out by teams with expertise in fields applied to the terrestrial environment and which occasionally translate the usual framework of their research to planetary science).

*3.3. The omnipresence of the Martian theme*

The study of the distribution of planetary science keywords related to Selection 2 (Fig. 4) shows the potential of the 3 study regions (Puna, Altiplano, and Atacama Desert) as a Martian analog. Work on meteorites, regolith and astrobiology also contributes significantly to the potential of these 3 regions in terms of planetary science. A nuance must however be brought on regolith because the selected articles involving this term do not necessarily have a direct link with planetary science. Such themes as Martian analogs, meteorites and astrobiology, which have a strong local contribution in South America, will be discussed later in this article.

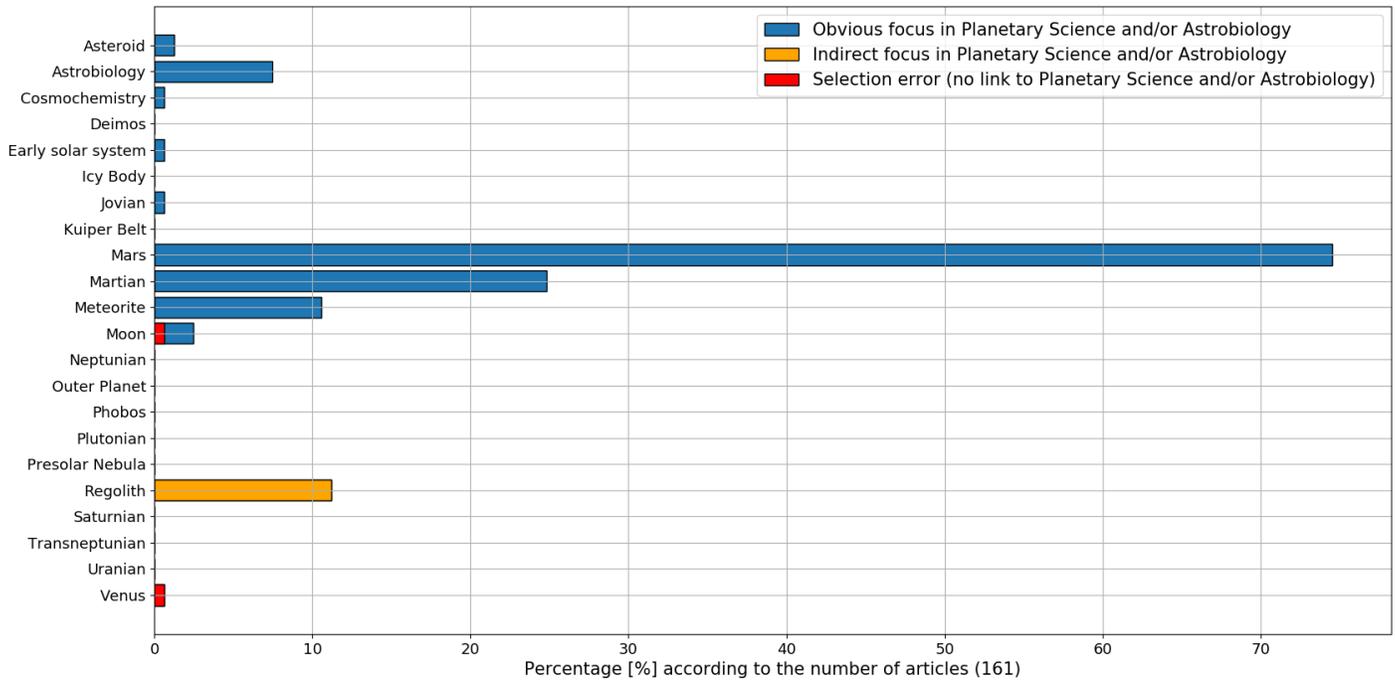

*Fig. 4 WoS data: Contribution of each keywords used to get the Planetary Science selection (i.e. Selection 2)*

## 4. Discussion

The main result of this study is that 60% of the work in planetary science based, in one way or another, on the exceptional environmental characteristics of the Puna region and the Atacama Desert do not included any local scientists (Table 5). Despite proven competences in geoscience, the Argentinean, Bolivian, Chilean, and Peruvian scientific communities have not yet succeeded in creating a specific study framework for planetary science.

*4.1. The sound basis of geoscience in South America*

This can refer as much to an ethical problem of international collaboration involving local partners (Budowski, 1975; Goss, 2021) as to a necessary introspection of the local scientific communities to understand their difficulties in scientifically enhancing their natural heritage and, above all, to protect and preserve it. The skills exist, as the statistics related to Selection 1 prove (Fig. 1 & Table 5). According to geoscience studies (i.e. Selection 1) research framework remains the same for both external and territorial countries: Argentina, Bolivia, Chile, and Peru having probably benefited from the technical progress linked to natural resources exploitation to structure an efficient academic environment. The University of Atacama, for example, located in the city of Copiapó was originally created to train mining technicians and engineers. A little further north, in the city of Antofagasta, the North Catholic University and the University of Antofagasta are historically linked to the exploration and exploitation of saltpeter and metallic ores as copper, silver and gold. Examples could be multiplied outside Chile for instance in Argentina with the National University of Catamarca, in Bolivia with the Technical University of Oruro or in Peru with the National University of San Augustin in Arequipa. This historical mining field has gradually been joined by the themes of biology and environmental conservation. In these fields, the South American countries mentioned above have a structured research and academic training system, which is largely lacking for the planetary science component. As previously mentioned, the multidisciplinary nature of this field of research does not facilitate the emergence of entities exclusively dedicated to planetary science, often associated with astronomy, geophysics, geochemistry, or biology. Furthermore, there is a notable difference in infrastructure between the external countries and the territorial countries. The United States hold the Lunar and Planetary

Institute (LPI) based in Houston, which was created in 1968 to give a scientific dimension to the Apollo program. The LPI also organizes one of the leading planetary science conferences each year, the Lunar and Planetary Science Conference (LPSC). In Europe, the countries present in Fig. 2 and Fig. 3 have at least one national institute (and often several) specialized in planetary science and, through the recent creation of the Europlanet society (launched in 2018 but with the first drafts of coordination dating back to the early 2000's), there is a desire to create a European synergy around this all too often fragmented theme of planetary science.

*4.2. The need to structure the field of planetary science in South America*

In South America, and more generally in Latin America, the fragmentation and lack of coordination in the field of planetary science seems to be widespread, even if a dynamic seems to be gradually being put in place, notably through the topic of astrobiology, which benefited from a Peruvian impetus in 2016 with the organization of the first Latin American Astrobiology Congress planned to be organized every two years and structured around a very active Mexican pole, stemming from the pioneering works of Dr. Rafael Navarro-González (Navarro-González et al., 2010). To the authors'knowledge, there are only two institutes in 2021 that openly display a planetary science component without being subordinated to any other dominant theme: in Cuba (Laboratorio de Ciencia Planetaria, Universidad Central "Marta Abreu" de Las Villas) and in Colombia (Grupo de Ciencias Planetarias y Astrobiología GCPA, Universidad Nacional de Colombia). However other initiatives are emerging, such as the organization of regional seminars to promote the use of data, often freely available, from space missions or civil society initiatives to promote space exploration. The absence of large-scale space programs in Latin American countries may be one of the reasons for the difference in the structure of planetary science with Europe and the USA (both having ambitious space mission programs). This diagnosis is reinforced by the preponderance of the NASA in the institutions linked to the publications selected via Selection 2 (Fig. 5). It is also interesting to note that China and in a lesser extend India, nations with strong space ambitions, have academic structures specialized in planetary science (Rong et al., 2021; Wei et al., 2018; Suzuki, 2019).

Of the four countries geographically hosting the Puna, Chile has the highest activity in planetary science. The presence of the world largest international astronomical observatories, which has surely favored the emergence of a structured community of Chilean astronomers, might also be related to this prevalence. Astronomy has indeed been the pioneer in terms of planetary science (Smith, 1997). But the link between the Chilean astronomical community and the geoscience community does not yet seem to be fully established, at least to develop structured research in planetary science.

Therefore, structuring this field of research on a Latin American scale, one based on pre-existing expertise in geoscience, should require the creation of a network of institutes, organizations and associations active and focused on planetary science and space exploration. Using this structure could help potential external scientists interested in working in this geographical area to increase the involvement of local partners by clearly identifying the correct interlocutors for their projects. It also seems necessary, beyond the scientific framework, to protect the remarkable environment of this exceptional natural heritage: for the purpose of environmental preservation but also to guarantee local scientific, technical and economic impact for each scientific study carried out in the region. In this respect, and in a way particularly focused on planetary science, the case of the Lanzarote and Chinijo Geopark in the Canary Islands may represent a relevant guideline (Mateo et al., 2019).

*4.3. Prospect of planetary science collaborations*

All things considered, a similar approach, combining environmental and planetary science studies, has been launched in 2021 in the Puna de Atacama (but based on preliminary studies from 2015), to assess the potential of this geographical region as a Martian analog, Mars being the focus of studies conducted in the Central Andes and related to planetary science (Fig. 4). This project, led by the University of Atacama in Chile, will be based on the implementation of a multidisciplinary high-altitude laboratory (Tavernier et al., 2020; Tavernier et al., 2021a) and the organization of joint campaigns with EMM program (Tavernier et al., 2021b; Tavernier et al., 2021c). The present study is part of a more general bibliographic inventory of high places of scientific interest in the Puna de Atacama and the Atacama Desert, which will be completed by fieldwork and remote sensing (Tavernier et al., 2021d; Medina et al., 2022). The ambition of such a project is to constitute, regionally and in the long term, a center of competence in planetary science open to the reception of international expertise and which could highlight, both locally and internationally, the exceptional natural heritage of the Central Andes.

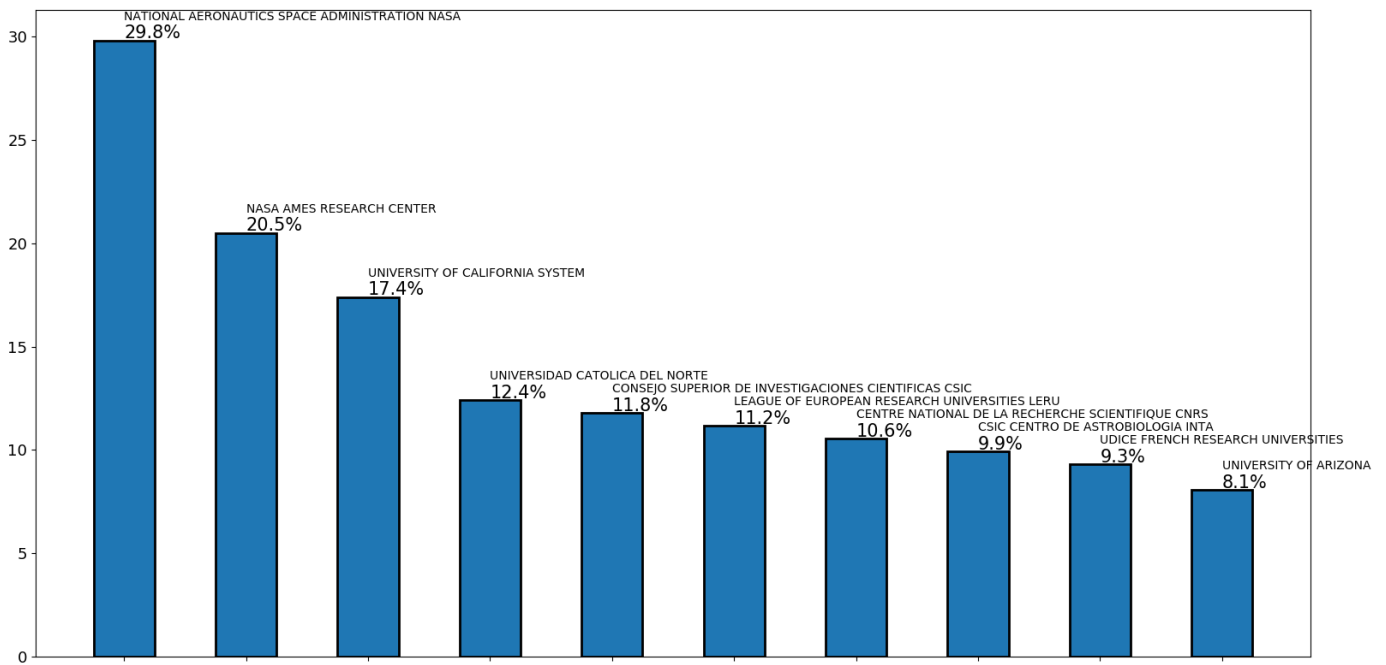

*Fig. 5 → WoS data: Contribution of different Institution to the Planetary Science selection (i.e. Selection 2)*

Meteoritics also represents, as is astrobiology in Mexico (De la Cruz et al., 2020), a base on the development of planetary science in Chile and Latin America. For instance, the analysis of chilean meteorites has been focused on different aspects such as (i) Distribution and antiquity (Gattacceca et al., 2011; Hutzler et al., 2016; Drouard et al., 2019), (ii) Solar System and planet formation (Pinto et al., 2021), (iii) Education (Pinto et al., 2020), and (iv) Protection of natural heritage (Valenzuela and Benado, 2018). The emergence of meteoritics in Chile appears related to the increased number of found meteorites in hot deserts (Scherer and Delisle, 1992; Bevan, 2006). The aridity and stable surfaces of the Atacama Desert provide exceptional characteristics to host areas with more than 150 meteorites/km²: one of the highest meteorite densities in the world (Hutzler et al., 2016). Multidisciplinary work in collaboration with international partners has enabled this field of study to be structured and to offer specific training for undergraduate and postgraduate Chilean students. In this sense, a new generation of geoscientists with meteorite-orientated skills is currently being developed, strengthening future projects for classifying and conserving meteorites in national repositories. A draft law is also being drawn up (in 2021) to protect this natural heritage. However, there is still a significant gap between (i) the field of classical geology, which is the driving force behind the mining industry in Chile, and (ii) observational astronomy and astrophysics, through the presence of the largest international observatories within the country. Despite this gap, meteorite research is gradually taking shape in Chile, as shown by the Chilean All-sky Camera Network for Astro-geosciences project (CHACANA) (Valenzuela et al., 2017). This project aimed to install four fireball detection cameras in the Antofagasta region but ultimately served as a pilot study held in Las Campanas and El Sauce observatories in northern Chile to provide a baseline for the global roll-out of the Fireball Recovery and InterPlanetary Observation Network (FRIPON) (Colas et al., 2020). This international network will allow monitoring of the skies to search for meteors, eventually leading to the recovery of fresh extraterrestrial samples that will help to raise the meteoritics discipline in Chile and South America.

However, these particular projects in Chile must be part of a wider perspective of structuring relevant research in planetary science, both on a national, continental, and worldwide scale.

**Acknowledgments**

The authors acknowledge Amy Marquardt for a careful review of the manuscript.

**Author contributions**

A. Tavernier designed the study, carried out the literature review, analyzed the data and wrote the paper, G.A. Pinto and M. Valenzuela wrote the section dealing with the meteorite research field in Chile, G.A. Pinto, M. Valenzuela, A. Garcia, C. Ulloa, R. Oses and B.H. Foing participated in the restructuring of the paper.

**Funding**

This study has been possible thanks to the Chilean Plan de Fortalecimiento de Universidades Estatales Año 2017 (ATA1799) and Año 2020 (ATA20992).

**Declaration of competing interests**

The authors declare that there is no conflict of interest regarding the publication of this article.

**Data availability**

The data is based on a keyword analysis of the WoS (https://www.webofscience.com/wos/woscc/basic-search), Scopus (https://www.scopus.com/search/form.uri?display=basic#basic) and ADS (https://ui.adsabs.harvard.edu/) websites. By following the methodology detailed in this article, it is possible to retrieve the results presented in this study. The data as well as the Python codes used to carry out the statistical study or the graphs of this article are available at the following Github directory https://github.com/ATavernierUDA/PunaAtacamaTrend

# Supplementary Materials

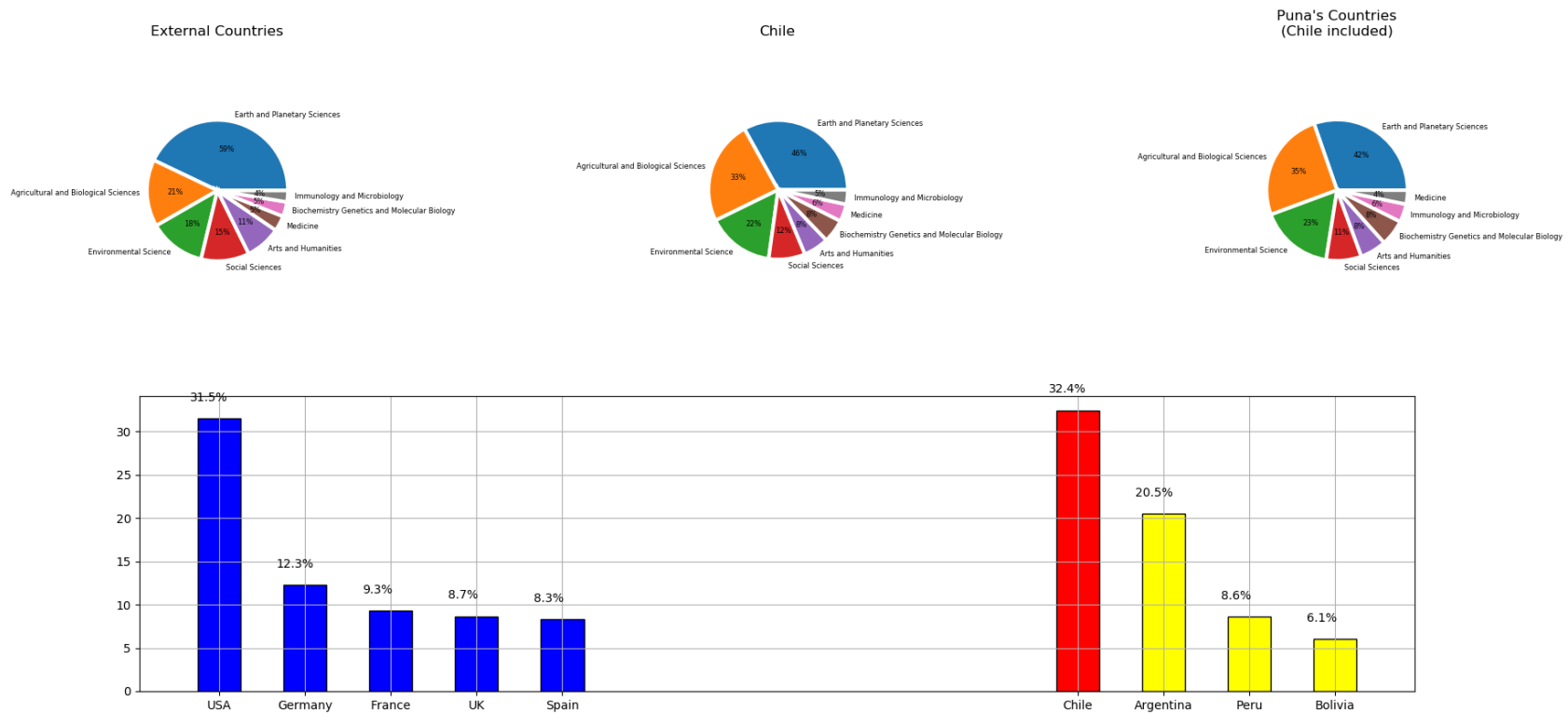

Fig. S1 Scopus Data: Relative contribution (according to the total number of publications) to the natural heritage of the Puna Region, in terms of publications, of different countries or group of countries (Selection 1, 5288 articles in total, excluding duplicates) | From top to bottom : distribution of field of research / percentage of contribution per country | Note: the fields of research are different from those of WoS (the use of an Earth and Planetary Science category does not allow a distinction between the different components of geoscience)

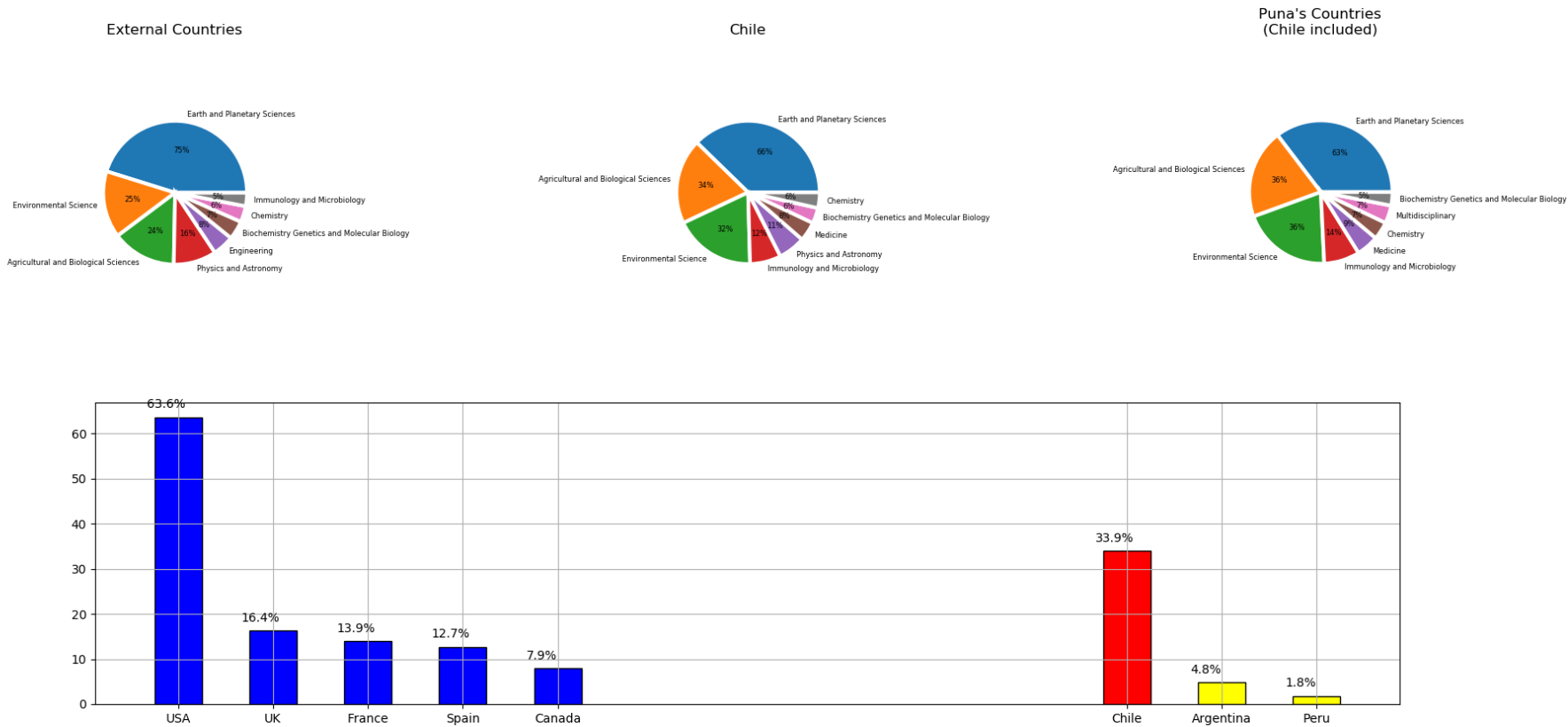

*Fig. S2 Scopus Data: Relative contribution (according to the total number of publications) to the natural heritage of the Puna Region, in terms of publications in Planetary Science, of different countries or group of countries (Selection 2, 165 articles in total, excluding duplicates) | From top to bottom : distribution of field of research / percentage of contribution per country | Note: the fields of research are different from those of WoS (the use of an Earth and Planetary Science category does not allow a distinction between the different components of geoscience)*

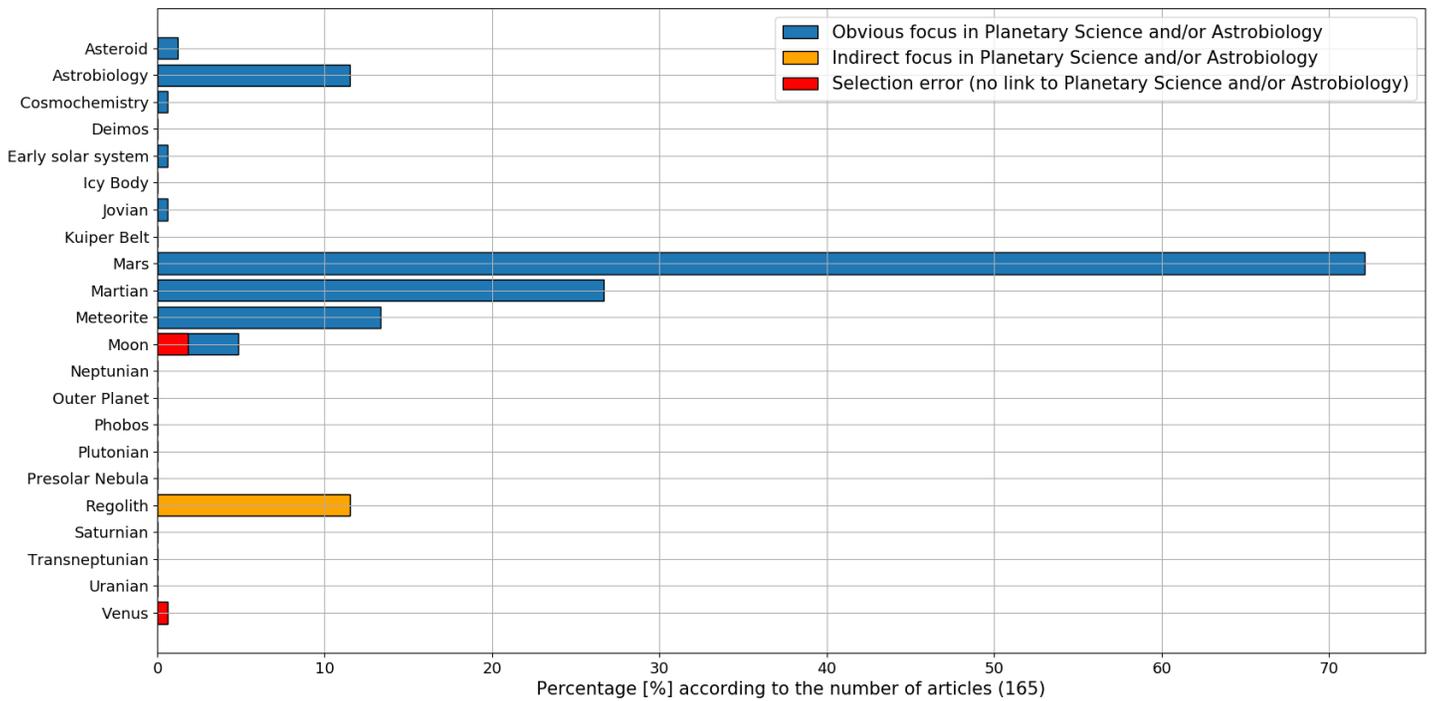

*Fig. S3 Scopus Data: Contribution of each keywords used in order to get the Planetary Science selection (i.e. Selection 2)*

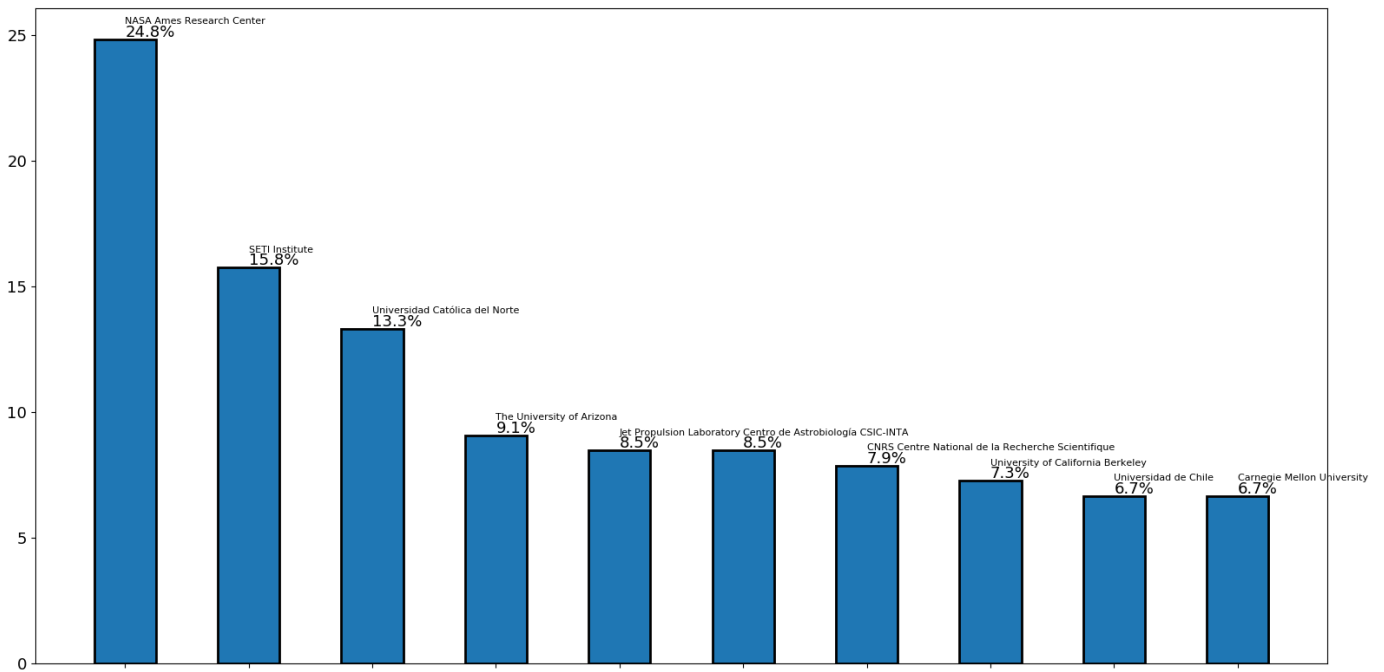

*Fig. S4 Scopus data: Contribution of different Institution to the Planetary Science selection (i.e. Selection 2)*

|  | Number of articles (excluding duplicates) | | |
| --- | --- | --- | --- |
|  | Total | External countries only | Participation of at least one territorial country |
| Selection 1 | 5288 | 1999 (38%) | 3289 (62%) |
| Selection 2 | 165 | 100 (61%) | 65 (39%) |

*Table S5: Overall contribution to Selection 1 & 2 from external and territorial countries (Scopus data)*